\documentclass{aa}

\usepackage{graphicx}
\usepackage{txfonts}
\begin{document}
\title{Fine-structure in the nonthermal X-ray emission of \\
SNR RX J1713.7$-$3946 revealed by Chandra}

\author{Yasunobu Uchiyama\inst{1}
          \and
          Felix A. Aharonian\inst{2}
          \and
          Tadayuki Takahashi\inst{1}
          }

\authorrunning{Y. Uchiyama et al.}

   \offprints{Yasunobu Uchiyama, \\ \email{uchiyama@astro.isas.ac.jp}}

   \institute{Institute of Space and Astronautical Science,
3-1-1 Yoshinodai, Sagamihara, Kanagawa 229-8510, and \\
Department of Physics, University of Tokyo, 
7-3-1 Hongo, Bunkyo, Tokyo 113-0033 \\
              \email{uchiyama@astro.isas.ac.jp; takahasi@astro.isas.ac.jp}
         \and
        Max-Planck-Institut f\"{u}r Kernphysik, Postfach 103980,
        D-69029 Heidelberg, Germany \\
              \email{Felix.Aharonian@mpi-hd.mpg.de}
             }

   \date{Received September 10, 2002; accepted December 9, 2002}

   \abstract{
We present morphological and spectroscopic studies  of
the northwest rim of the supernova remnant RX~J1713.7$-$3946 
based on observations  by  the \emph{Chandra X-ray observatory}.
We found a complex network of nonthermal (synchrotron)
X-ray filaments, as well as a 'void' type structure
-- a dim region of a circular shape -- in the northwest rim.
It is remarkable that  despite distinct brightness variations, 
the X-ray spectra everywhere in this region can be well fitted with
a power-law model with photon index ranging $\Gamma =$ 2.1--2.5. 
We briefly discuss some  implications of these results 
and argue  that the resolved  X-ray features in  the  
northwest rim may challenge the perceptions  of  
the standard diffusive shock-acceleration models 
concerning the  production, propagation and radiation  
of relativistic  particles in supernova remnants.

   \keywords{radiation mechanisms: non-thermal --
             ISM: supernova remnants --
             ISM: cosmic rays --
             stars: supernovae: individual: RX J1713.7$-$3946
               }
   }

\titlerunning{X-ray fine-structure of SNR RX J1713.7$-$3946}

   \maketitle

%________________________________________________________________
\section{Introduction}

The supernova remnant (SNR) RX~J1713.7$-$3946
has proven to be a prominent source of nonthermal X-rays 
and presumably also $\gamma$-rays, thus providing 
strong evidence that  shell-type SNRs are sites of acceleration 
of galactic cosmic rays.
This source was  discovered during the \emph{ROSAT} 
All-Sky Survey (Pfeffermann \& Aschenbach \cite{pfeffermann}). 
Observations with \textit{ASCA} have revealed intense
synchrotron X-ray emission from the entire
remnant (Koyama et al.\ \cite{koyama97}; Slane et al.\ \cite{slane}).
Unlike SN~1006, no evidence for thermal X-ray components has yet been found.

At the north perimeter of RX~J1713.7$-$3946,
a molecular cloud (cloud~A) has been  found with a high
CO\,($J$=2--1)/CO\,($J$=1--0) ratio that suggests possible  
interaction between  the cloud and the SNR shell
(Butt et al.\ \cite{butt01}). The cloud has positional association with 
the unidentified $\gamma$-ray source 3EG J1714$-$3857 
(Butt et al.\ \cite{butt01}).
Most recently, Uchiyama et al.\ (\cite{uchiyama02})  reported
an unusually flat-spectrum X-ray source (AX J1714.1$-$3912)
coincident with this cloud, and  argued that the flat spectrum can be 
best interpreted by bremsstrahlung from either sub-relativistic protons 
or mildly relativistic electrons. 
The high-energy particles responsible for the X-ray and $\gamma$-ray emission 
from this cloud are likely associated (in one way or another) 
with the SNR-cloud  interaction.  
For the preferred distance to SNR RX~J1713.7$-$3946 of $d \simeq 6$~kpc
(Slane et al.\ \cite{slane}),
based on the kinematic distance to cloud~A, the age of the remnant 
is estimated to be $\gtrsim 10 \,000$~yr.  
A younger age of $\sim 2000$~yr 
cannot be, however, excluded if $d=\mbox{1--2}$~kpc.

The CANGAROO collaboration (Muraishi et al.\ 2000) reported  
the detection of TeV $\gamma$-ray emission from the direction 
of the northwest (NW) rim,  the brightest region of synchrotron X-rays 
in RX~J1713.7$-$3946. If confirmed, the TeV $\gamma$-radiation 
would provide direct and  unambiguous evidence for the presence 
of particles (electrons and/or protons)  accelerated to very high 
energies.  Recently Enomoto et al.\ (\cite{enomoto}) published   
the spectrum of TeV emission based on the new CANGAROO 
observations. The spectrum  is claimed to be quite  steep
with  a  power-law photon index 
$\Gamma = 2.8\pm 0.2$ between 400 GeV and 8 TeV. They  
argued that the steep spectrum  is inconsistent with the 
inverse  Compton (IC) model,  but could be explained 
by $\pi^0$-decay gamma-rays. If true,  for the canonical 
shock-acceleration spectrum of protons with power-law 
index $s \sim 2$, this would imply  a cutoff energy in 
the proton spectrum around 100 TeV. Subsequently, 
Reimer \& Pohl (\cite{reimer}) and Butt et al.\ (\cite{butt02}) 
argued that this interpretation would violate 
the $\gamma$-ray flux upper limits set by the EGRET instrument.
However, this is not a sufficiently robust argument  to be used 
to dismiss the hadronic origin of the reported TeV emission. 
Adopting a slightly harder proton spectrum, 
e.g. with spectral index $s \leq 1.9$, 
it is possible to avoid the conflict with the EGRET data.
Even for proton spectra steeper than $s=2$, 
it is still possible to suppress the GeV $\gamma$-ray flux, 
if one invokes the effects of energy-dependent propagation  
of protons while traveling from the  accelerator (SNR shocks) 
to the nearby clouds (see e.g. Aharonian \cite{aharonian01}).
Moreover, the lack of GeV $\gamma$-rays can be naturally explained by 
confinement of low-energy (GeV)  protons in the supernova shell, 
in contrast to the effective  escape of high-energy (TeV) protons.

On the other hand, the arguments against the IC model of TeV emission 
should be backed by thorough theoretical studies based on higher quality 
data from the radio, X-ray and $\gamma$-ray domains.
X-ray observations are of particular interest   because the synchrotron X-ray 
spectra reliably ``control''  the predictions of IC emission at TeV energies.
In this paper, we report on the X-ray study of 
the NW rim of SNR RX~J1713.7$-$3946 
using archival data obtained with the \emph{Chandra X-ray Observatory} 
(Weisskopf et al.\  \cite{weisskopf96}). 
We demonstrate that the X-ray emission
has remarkable substructure with  bright filaments 
accompanied by curious dark voids.
The observed features set new standards in X-ray studies of 
SNRs which should help to understand deeper  the nonthermal 
processes  of particle acceleration,  propagation and radiation 
in supernova shocks.

%_____________________________________________________________
\section{X-ray data}

\begin{figure}
   \centering
    \includegraphics[width=6cm]{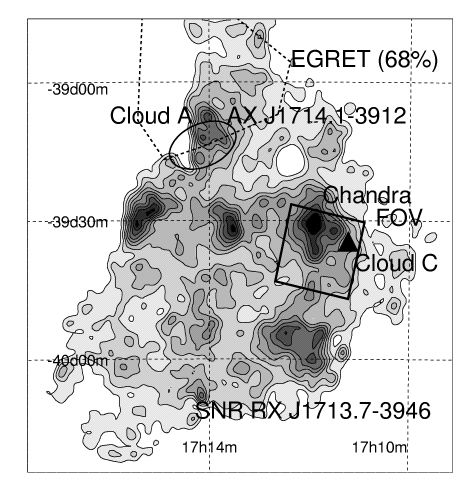}
      \caption{
Overall \emph{ASCA} view of the SNR RX~J1713.7$-$3946 region
in the hard 5--10 keV energy band (Uchiyama et al.\ \cite{uchiyama02}), and
the \emph{Chandra}'s $17\arcmin \times 17\arcmin$ field-of-view.
The SNR has an angular radius of about $40\arcmin$.
A transient X-ray source (Koyama et al.\ \cite{koyama97})
is blanked out.
Depicted are the 68\% confidence error contour of the EGRET 
unidentified source 3EG J1714$-$3857 (Butt et al.\ \cite{butt01}),
the approximate outline of the cloud~A, 
and the center of the cloud~C (\emph{a triangle}).
}
         \label{fig:fov}
   \end{figure}

The observation of the northwest (NW) rim of the SNR RX~J1713.7$-$3946 
was carried out on 2000 July 25 (ObsId 736)
with the Advanced CCD Imaging Spectrometer (ACIS)
on board \emph{Chandra}.
The data were obtained with the ACIS-I array 
consisting of four front-side-illuminated CCD chips.
As shown in Fig.\,\ref{fig:fov},  
the array was placed to cover the brightest portion of the remnant.
We analyzed the processed level 2 event data 
(processing version R4CU5UPD14.1), 
retrieved from the \emph{Chandra} public archive,
with a new gain map applied.
The overall light-curve was examined for possible 
background flares or fluctuations;  no  significant time-variation was found. 
During such a quiescent period, the instrumental background is 
relatively constant ($\pm 10\,\%$) with time 
\footnote{For details, see http://cxc.harvard.edu/contrib/maxim/bg/ } .
Indeed, the count rate in the 8.5--10 keV energy interval is typically 
only several percent lower than the (standard) blank-sky data sets.
Since this difference is within the statistical uncertainties of our results 
presented in this paper,
the instrumental background component is taken from the blank sky data.
After data  processing, we obtained an effective exposure time of 29.6 ks.

%_____________________________________________________________
%                 A figure as large as the width of the column
%-------------------------------------------------------------
%

%_____________________________________________________________
\subsection{Image analysis}

\begin{figure}
   \centering
     \includegraphics[width=6cm]{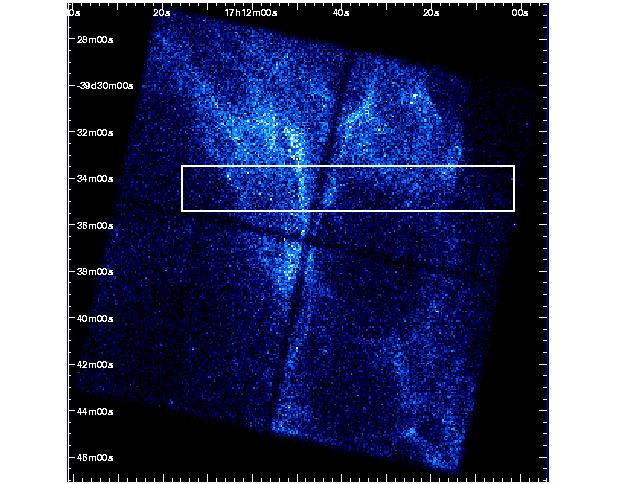}
      \caption{\emph{Chandra} ACIS-I image of 
the NW rim of RX~J1713.7$-$3946 in the broad energy band (1--5~keV)
with the photon counts (linear scale from 0 to 15 counts) 
accumulated in $8\times8$ pixels. 
The coordinates are R.A. and decl. (J2000).
}

         \label{fig:image8}
   \end{figure}

\begin{figure}
   \centering
      \includegraphics[width=8cm]{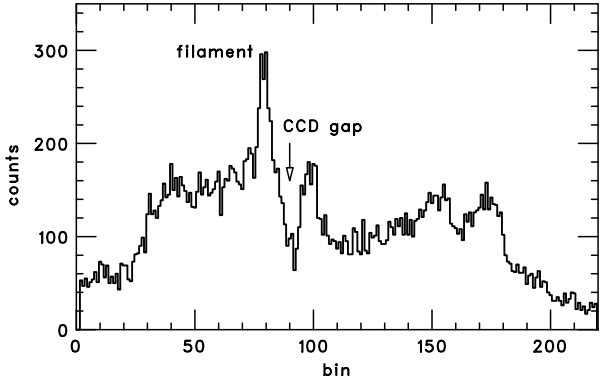}
      \caption{Horizontal profile of the photon counts (1--5 keV) for the rectangle
region shown in Fig.\,\ref{fig:image8}, integrated over the vertical dimension.
Distance is in units of $4\arcsec$, from east to west.}
         \label{fig:belt}
   \end{figure}

\begin{figure}
   \centering
     \includegraphics[width=6cm]{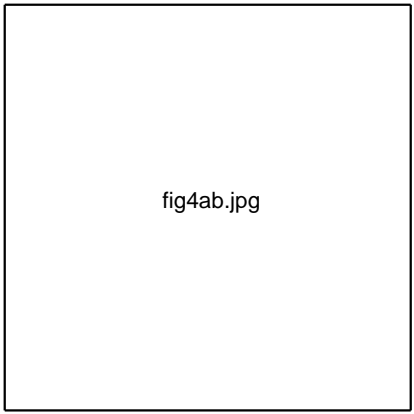}
      \caption{Smoothed image of the NW rim of RX~J1713.7$-$3946 
in \textbf{a)} soft 1--3 keV band and \textbf{b)} hard 3--5 keV band, 
obtained with four (each with $8\farcm 4 \times 8\farcm 4$) CCD chips.
The scales are in units of counts s$^{-1}$ arcmin$^{-2}$.
The corrections for the inter-chip gaps and vignetting have not been applied.
}
         \label{fig:ximage}
   \end{figure}

\begin{figure}
   \centering
     \includegraphics[width=6cm]{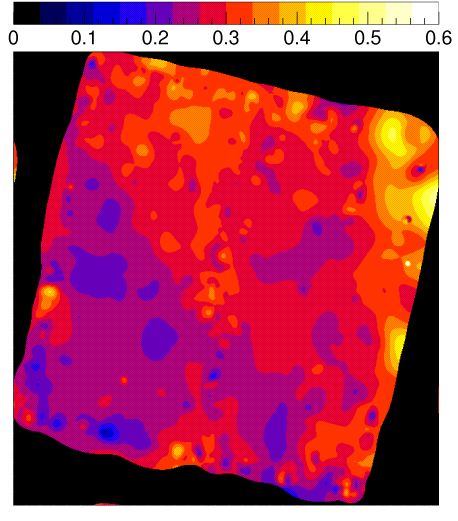}
      \caption{The spatial distribution of the hardness ratio, 
(3--5 keV)/(1--3 keV), in  the NW rim, obtained by 
dividing the smoothed brightness map 
(in units of $\mathrm{photons}\ \mathrm{s}^{-1}\, \mathrm{arcmin}^{-2}$)
in the 3--5 keV energy band to  that in the 1--3 keV band.}
         \label{fig:hardness}
   \end{figure}

Figure \ref{fig:image8} shows the \emph{Chandra} image of the NW rim of 
SNR RX~J1713.7$-$3946 in the energy range of 1--5 keV with
the photon counts summed in pixels of $4\arcsec\times4\arcsec$. 
Point sources are removed, based on a wavelet decomposition 
using the CIAO software \textsf{wavdetect} (Freeman et al.\ \cite{freeman02}).
We extracted the profile of photon counts from the rectangular region 
depicted in Fig.\,\ref{fig:image8}, summed over the vertical 
narrow dimension of the rectangle (Fig.\,\ref{fig:belt}).
Figure \ref{fig:ximage} shows the smoothed X-ray images in the soft 1--3 keV 
and hard 3--5 keV bands.
The images have been adaptively smoothed with the CIAO software
\textsf{csmooth}, where the local photon counts are utilized to determine the
width of the Gaussian smoothing kernel at each position.
The map of kernel widths was calculated based on the broad 1--5 keV image.
In Fig.\,\ref{fig:hardness}, we show the spatial distribution of the 
hardness ratio, (3--5 keV)/(1--3 keV),  of the smoothed images, after corrections 
for the mirror vignetting and  exposure time.

%_____________________________________________________________
%                 A figure as large as the width of the column
%-------------------------------------------------------------
%

%_____________________________________________________________
%                 A figure as large as the width of the column
%-------------------------------------------------------------
%

%_____________________________________________________________
%                 A figure as large as the width of the column
%-------------------------------------------------------------
%

%_____________________________________________________________
%                 A figure as large as the width of the column
%-------------------------------------------------------------

Dozens of compact and  extended emission features  have been detected 
throughout this region.
The \emph{ASCA} image of the NW rim now is resolved 
to distinctive components.  Chains of bright \emph{filaments} 
are evident within a more diffuse and fainter \emph{plateau}.
The surface brightness of the filaments is on average about two times  higher 
than that of the surrounding plateau, although the latter 
constitutes the bulk ($\sim 80$\,\%) of 
the X-ray flux detected from the rim region comprised of the filaments and
plateau.
The most prominent and largest filament  located near the center of the image and 
running from north to south, 
has an apparent width of $\theta \simeq 20\arcsec$,
which corresponds to linear  scale of 
$\Delta R \simeq 0.58\,d_6$ pc,
where $d_6$ is the distance in units of 6 kpc.
Adjacent to the west side of the prominent filament,
there can be seen an interesting morphology -- a  ``\emph{void}'',  
i.e.  a dim circular region with a radius of 3\arcmin .
South of this circular void,  we can see another void structure 
which  seems  to be extending 
beyond the observed field.

The features mentioned above are observed both in the soft and hard images.
The two images are strikingly similar, thus implying small spectral variations
across the NW rim including the filament, plateau, and dark void regions.
This is quantitatively shown in the hardness-ratio map (Fig.\,\ref{fig:hardness}).
The northwest corner of the observed field, outside the boundary of 
this SNR,  shows relatively hard emission, which would 
originate in hot optically-thin thermal plasmas pervading the Galactic plane.

%_____________________________________________________________
\subsection{Spectral analysis}

\begin{figure}
   \centering
     \includegraphics[width=6cm]{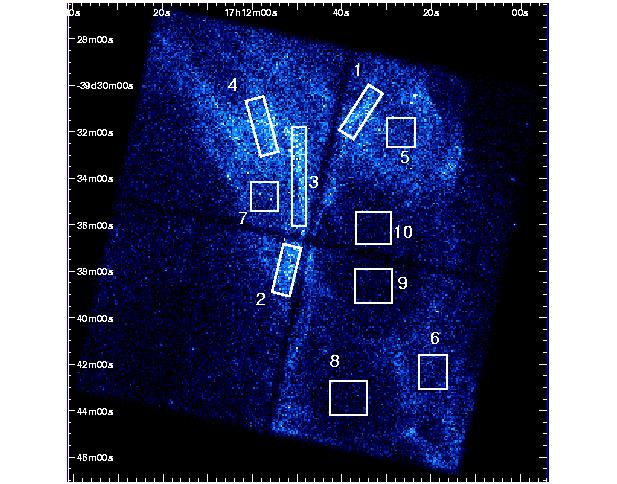}
      \caption{The regions selected  for  the spectral analysis, 
overlaying the same image as Fig.\,\ref{fig:image8}. 
The rectangular regions were selected for the bright filaments (1--4),
plateau (5--7), and dark voids (8--10).
}
         \label{fig:label}
   \end{figure}

We have examined  the  X-ray spectra extracted from various 
regions  in the NW rim, including the filaments, plateau, and void 
regions (Fig.\,\ref{fig:label}).
The instrumental background components
are subtracted by using blank-sky observations. 
Since  the contribution of diffuse Galactic X-ray emission 
to the spectra is estimated to be about 5\,\% and 10\,\% for the 
filament and plateau regions, respectively, 
it has negligible effects on the spectral properties derived from these regions.
Note, however, that for the void regions, the contamination could go up to 20\,\%.

%_____________________________________________________________
%                 A figure as large as the width of the column
%-------------------------------------------------------------

\begin{figure}
   \centering
       \includegraphics[width=8cm]{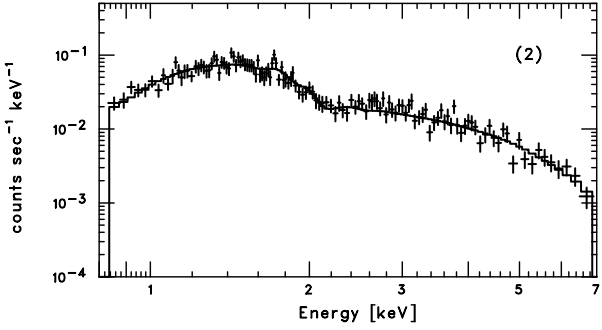}
        \caption{ACIS-I spectrum of the filament region 2.
Superposed on the data points is histogram of the best-fit power-law model
with interstellar absorption.}
         \label{fig:spec}
   \end{figure}

\begin{figure}
   \centering
        \includegraphics[width=8cm]{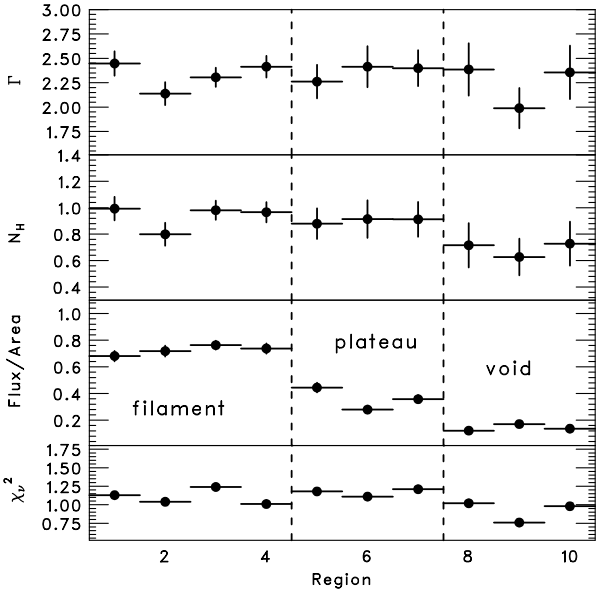}
       \caption{Results of spectral fits with an absorbed power-law 
model for several selected  regions: the bright filaments (region 1--4),
plateau (region 5--7), and dark voids (region 8--10).
From top to bottom panels, plotted are
the best-fit values (with their 90\,\% errors) of photon index $\Gamma$ and
absorbing column density 
$N_{\rm H}$ in units of $10^{22}\ \rm cm^{-2}$,
surface brightness in units of $10^{-12}\, \mathrm{erg}\ \mathrm{cm}^{-2}\,
\mathrm{s}^{-1}\, \mathrm{arcmin}^{-2}$, and reduced chi-square $\chi_\nu^2$.
}
         \label{fig:fit_result}
   \end{figure}

Like the spatially integrated spectrum that has been 
discovered by \emph{ASCA}, the spectra from individual 
regions  can be represented as  \emph{featureless continuum}.
This can be best interpreted by synchrotron emission produced 
by relativistic electrons with multi-TeV energies. 
We fit the spectrum in the 0.8--7 keV interval
individually with a power-law model taking account of photoelectric 
absorption along the line-of-sight, and obtain statistically acceptable fits in all cases. 
In Fig.\,\ref{fig:spec}, we present the energy spectrum extracted from 
region 2 as a characteristic  example, together with the best-fit model.
We show the spectral parameters for these regions
in Fig.\,\ref{fig:fit_result}.  It is  demonstrated 
that the best-fit values of photon index $\Gamma$
[$F(\varepsilon)=K\varepsilon ^{-\Gamma}$ 
(photons cm$^{-2}$ s$^{-1}$ keV$^{-1}$)] and absorbing column density 
$N_{\rm H}$ are remarkably similar to each other, despite substantial 
brightness  variations, with average values  of $\Gamma \simeq 2.3$ and 
$N_{\rm H} \simeq 0.9 \times 10^{22}\ \rm cm^{-2}$.
These parameters agree fairly well with the reported  
\emph{ASCA} spectrum integrated over the NW rim:  
$\Gamma = 2.4\pm 0.1$ and 
$N_{\rm H} \simeq (0.81\pm 0.06) \times 10^{22}\ \rm cm^{-2}$
(Koyama et al.\ \cite{koyama97}).

%_____________________________________________________________
%                 A figure as large as the width of the column
%-------------------------------------------------------------
%

%_____________________________________________________________
%                 A figure as large as the width of the column
%-------------------------------------------------------------

Generally,  a simple power-law function provides
very convenient and meaningful parametrizations, especially 
for nonthermal X-ray in limited energy intervals.
However,  the synchrotron origin of X-ray emission {\em a priori} 
implies the presence of spectral cutoffs associated with the unavoidable 
(see below)  cutoff  in the acceleration spectrum of parent electrons. 
More specifically, in the case of SNRs the spectrum of synchrotron X-rays 
is formed by the highest-energy electrons distributed in the cutoff region.  
Although, the current models  of diffusive shock acceleration  do not provide definite
predictions about the spectral form of electron distribution in the cutoff region,
an  {\em exponential cutoff}  is commonly believed to be a natural 
assumption. Below we will adopt a more relaxed (generalized) form for 
the electron distribution, namely  
%eq1
\begin{equation}
\label{electrons}
N(E)  \propto  E^{-s} \exp \left[ - \left(\frac{E}{E_{\rm m}} \right) ^{\beta} \right ] \ , 
\end{equation}
with $\beta=$1/2, 1,  and 2, where $E_\mathrm{m}$ is the maximum energy 
of accelerated electrons.

For this energy distribution of electrons, 
the  $\delta$-functional approximation gives a simple analytical 
presentation for the 
differential spectrum of synchrotron radiation:
%eq2
\begin{equation}
\label{photons}
F(\varepsilon)  \propto  \varepsilon ^{-\Gamma} 
\exp \left[ -\left( \frac{\varepsilon}{\varepsilon_0} \right) ^{a} \right],
\end{equation}
where   $a=\beta/2$, $\Gamma=(s+1)/2$ is the photon index and 
$\varepsilon_0$ is the cutoff energy. Generally,
$\varepsilon_0=\zeta \varepsilon_{\rm c}$, where  
%eq3
\begin{equation}
\label{synchenergy}
\varepsilon_{\rm c} \simeq 5.3 \left (\frac{B}{10 \ \mu \rm G} \right) 
\left(\frac{E_{\rm m}}{100 \ \rm TeV} \right )^2 \ \rm keV \ ,
\end{equation}
is the characteristic energy of synchrotron photons emitted by an electron of energy 
$E=E_{\rm m}$, and $\zeta$ is a parameter introduced  to adapt 
this presentation to the results of the accurate numerical calculations. 
These calculations  show that  for $\zeta=1$ (but not for $\zeta  \simeq 0.3$ 
as often used in literature),   
Eq.\,({\ref{photons}) gives an  adequate description of the  
synchrotron spectrum with accuracy $\leq 25\, \%$  in a rather broad energy band, 
including the cutoff region (e.g. Aharonian \cite{aharonian00}).   
 
Therefore below we invoke the spectral presentation 
given by Eq.\,({\ref{photons}) 
in order to estimate  the  allowed  range for the cutoff-energy position in the 
\emph{Chandra} data. We treated the  cutoff energy 
$\varepsilon_0$ as a free parameter, and derived the best-fit values 
for different combinations of fixed parameters  
$\Gamma$ and $a$, namely $\Gamma = 1.5, 1.75, 2.0$ and 
$a =1, 1/2, 1/4$. We note that for the standard shock-acceleration 
spectrum with $s=2$, one should expect $\Gamma=1.5$, provided 
that the spectrum of electrons is not suffered deformation due to 
radiative and non-radiative (e.g. escape) losses. 
However,  typically the cooling time of highest energy 
electrons responsible for synchrotron X-ray emission is   
shorter  than the source age, even for young, 1000 yr old SNRs. 
This leads  to the spectral steepening, 
thus,  $\Gamma \sim  2.0$ (corresponding to $s=3$) would be 
a more realistic choice. 

We have performed spectral fits for one of the filamentary regions,
region 2, by the model defined by Eq.\,({\ref{photons}) together with
interstellar absorption.
We have obtained good fits  ($\chi_\nu^2 \simeq 1.0$, $\nu = 127$) 
for each set of parameters, the   
absorbing column densities being found in the range
$N_{\rm H} =  \mbox{(0.66--0.79)} \times 10^{22}\ \rm cm^{-2}$.
The best-fit cutoff energies are presented in Table~\ref{tab:cutoff}.  
Except for $a=1/4$  (which actually implies very 
slow steepening of the electron spectrum rather than a cutoff), 
the synchrotron cutoff energies are high. In particular for  
the  most realistic case  with $\Gamma = 2.0$, 
the synchrotron cutoff energy exceeds 10 keV.

%_______________________ table
   \begin{table}
      \caption[]{Cutoff photon energy obtained for the synchrotron filament (region 2).}
         \label{tab:cutoff}
%{p{0.5\linewidth}l}
     $$ 
         \begin{array}{p{0.2\linewidth}p{0.2\linewidth}l}
            \hline
            \noalign{\smallskip}
            \multicolumn{2}{l}{\mathrm{Parameter}}  &  \varepsilon_0 \ (\mathrm{keV})\\
            \noalign{\smallskip}
            \hline
            \noalign{\smallskip}
            $\Gamma = 1.5$ &  $a = 1$     &  5.2_{-0.8}^{+1.1}    \\
                                        &  $a = 1/2$  &  2.0_{-0.6}^{+1.0}    \\
                                        &  $a = 1/4$  &  0.08_{-0.04}^{+0.09}    \\
            \noalign{\smallskip}
            $\Gamma = 1.75$ &  $a = 1$     &  8.4_{-1.9}^{+3.4}    \\
                                        &  $a = 1/2$  &  5.0_{-1.9}^{+5.6}    \\
                                        &  $a = 1/4$  &  0.44_{-0.26}^{+1.8}    \\
            \noalign{\smallskip}
            $\Gamma = 2.0$ &  $a = 1$     &  \geq 12.2     \\
                                        &  $a = 1/2$  &  \geq 11.7     \\
                                        &  $a = 1/4$  &  \geq  2.8     \\
%                                        &  $a = 1/2$  & \leq 1700^{\mathrm{a}}     \\
            \noalign{\smallskip}
            \hline
         \end{array}
     $$ 
%\begin{list}{}{}
%\item[$^{\mathrm{a}}$] This is foot a
%\end{list}
   \end{table}
%

%
%______________________________________________________________
\section{Discussion}

A striking result of this study  is the discovery of the  
inhomogeneous, filamentary  structure in the synchrotron X-ray emission
of the  NW rim.  The  bright \emph{filaments} and {\em hot spots}
embedded in the diffuse \emph{plateau} emission, cannot be 
easily explained  by  the standard diffusive shock
acceleration theory.  More surprisingly, 
the X-ray spectrum does not show noticeable 
variations  over the entire rim, in spite of  the 
strong gradient of the surface   brightness distribution.     

The derived hard power-law spectra of synchrotron X-rays
imply that the electron acceleration continues effectively 
to  $\geq 100$ TeV, unless we assume very strong magnetic field 
exceeding 100 $\rm \mu G$.  Even so, independent of the strength of 
magnetic field,  the standard  shock acceleration theory predicts a cutoff in the 
spectrum of synchrotron radiation around  $1 \ \rm keV$ or below, 
and correspondingly -- steep X-ray spectra. 
%Indeed,  the radiative energy losses of relativistic 
%electrons in the  environment with magnetic field exceeding 
%$3 \ \rm \mu G$ is dominated by synchrotron cooling (taking into account that
%the electron cooling on the diffuse galactic optical and infrared 
%radiation fields is significantly reduced due to the Klein-Nishina effect).     

In the framework of diffusive shock acceleration model, the synchrotron cutoff  
energy is set by the condition 
\emph{acceleration rate = synchrotron loss rate}.
The synchrotron cooling time of an electron of energy $E_{\rm e}$ 
in the ambient magnetic field of strength $B$ is
%eq4
\begin{equation}
\label{synchtime}
t_{\rm synch} =  1.25 \times 10^3 
\left ( \frac{E_{\rm e}}{100 \ \rm TeV} \right)^{-1} 
\left (\frac{B}{10 \rm \ \mu G} \right )^{-2}  \  \rm yr \  . 
\end{equation}    

The diffusive shock acceleration time 
(see e.g.  Malkov and Drury \cite{malkdrury01}) 
can be written, with accuracy of about $\pm 50\,\%$, in a simple form 
$t_{\rm  acc} =  10 \mathcal{D}/V^2$,
where $\mathcal{D}$  is the diffusion coefficient 
in the upstream region, and $V$ is the upstream velocity into the shock. 
The diffusion coefficient is generally 
highly unknown parameter, however, if one requires acceleration of 
particles to highest energies allowing synchrotron radiation up to X-ray 
energies, we must assume that the diffusion proceeds in the 
so-called Bohm diffusion regime, therefore it is convenient to 
parametrize the diffusion coefficient in terms of 
the Bohm diffusion coefficient 
$\mathcal{D}_{\rm B}=\eta r_{\rm g} c/3$, where $r_{\rm g}=E_{\rm e}/eB$ and 
$\eta \geq 1$  are  the so-called gyroradius and gyrofactor, respectively.  
The value $\eta=1$ implies  the smallest possible value of the 
diffusion coefficient, and correspondingly the shortest  possible acceleration time,
%
%eq5
\begin{equation}
\label{acctime}
t_{\rm acc}
\approx 
2.7 \times 10^3   
\left ( \frac{E_{\rm e}}{100 \ \rm TeV} \right) 
\left (\frac{B}{10 \rm \ \mu G} \right )^{-1}    
\left (\frac{V}{2000 \ \rm km\ s^{-1}}  \right )^{-2}  \eta  \  \rm yr \  . 
\end{equation}    
From Eqs.\,(\ref{synchtime}) and (\ref{acctime}) we find
the maximum energy of accelerated  electrons,
% eq6
\begin{equation}
\label{Emax}
E_{\rm m} \approx 67 
\left (\frac{B}{10 \rm \ \mu G} \right )^{-1/2}    
\left (\frac{V}{2000 \ \rm km\ s^{-1}}  \right ) \eta^{-1/2} \ \rm TeV \ .
\end{equation}    
Now, substituting $E_{\rm m}$ from Eq.\,(\ref{Emax}) into Eq.\,(\ref{synchenergy}),
we find that the synchrotron cutoff energy depends  only on the shock speed and 
the gyrofactor, 
\begin{equation}
\label{synchcutoff}
\varepsilon_0 \approx 2  \ 
\left (\frac{V}{2000 \ \rm km\ s^{-1}}  \right )^2\,  \eta^{-1} \ \rm keV \ .
\end{equation}    

This simple treatment does not tell us much how far the 
electron spectrum can be  extended  beyond  $E_{\rm m}$,  
and correspondingly how looks like the synchrotron spectrum 
in the cutoff region around  $\varepsilon_0$.  
The assumption that  the electron acceleration spectrum 
is described  by the ``canonical'' Eq.\,(\ref{electrons}) is  indeed 
a rough approximation.  
In fact,  the highest energy electrons from  the cutoff region may  
have more complicated distributions, which, for example, may  contain  a pronounced pile-up just proceeding the cutoff 
(e.g. Melrose \& Crouch \cite{melrose97}). 
However, this may only shift the position of the synchrotron cutoff to lower energies 
(Aharonian \cite{aharonian00}). 
Finally, we should note that  the position of the synchrotron 
cutoff given  by Eq.\,(\ref{synchcutoff}) should be considered as an absolute 
upper limit because it does not take into account other possible losses. 

The results in the previous section
indicate that the synchrotron cutoff in the radiation components
observed from the NW rim,   most likely are   
located  above  10  keV. This requires   
$V \gtrsim 5000\,\eta^{1/2}\ \mathrm{km\ s}^{-1}$
which for any reasonable shock speed
in RX~J1713.7$-$3946 is quite a tough condition  
and hardly could  be fulfilled  even when the acceleration 
proceeds in the extreme Bohm diffusion regime ($\eta = 1$).  
A possible solution to this difficulty could be 
found if one assumes that   the electron acceleration and  
radiation regions are  effectively  separated. Indeed,  assuming   that
the  acceleration takes   place in sites of  low-magnetic field,  
then the electrons  quickly  escape the  acceleration region, 
enter into  regions  with significantly
enhanced magnetic field, and  there produce the bulk of the observed 
X-ray flux  with  a  synchrotron cutoff  well above 1 keV.     
Otherwise, one should invoke faster (yet unknown) electron acceleration 
than the standard shock-acceleration model provides, 
in order to account for the observed hard synchrotron X-ray spectra.
Another possibility would be if the electrons are of secondary  
($\pi^\pm$-decay) origin produced at interactions of accelerated protons 
and ions with the ambient gas. This hypothesis 
requires very strong magnetic field and acceleration of protons 
to energies $\geq 10^{15} \ \rm eV$. Such a model  
recently was suggested by Bell \& Lucek (\cite{bell}).
This hypothesis implies very hard spectrum of $\pi^0$-decay TeV $\gamma$-rays
with energy flux comparable to the energy flux of synchrotron X-ray emission.
Therefore, the ``hadronic'' origin of electrons would be turned down, if the 
TeV $\gamma$-ray spectrum indeed breaks  
at sub-TeV energies as has been claimed by Enomoto et al.\ (\cite{enomoto}).

Formally, the X-ray brightness distribution  seen in  Fig. \ref{fig:ximage} 
could be result of  different depths of the line of sight, i.e. assuming that 
the bright filaments are sheets seen at ``edge on'', and the plateau 
emissions are those seen at ``face on''.
Below we adopt a more likely, in our view,  interpretation that 
the observed brightness distribution  is  a result of highly   
inhomogeneous production and distribution  of  relativistic electrons  and   
magnetic fields in the NW rim.  
At first glance,  homogeneous  production of  
multi-TeV electrons over the NW rim  cannot be
ruled out, because in this case  the filamentary structure   
might  be referred  to  a  local enhancement of the magnetic field. 
However, in the (most likely) regime when  the 
synchrotron cooling time is less than the source age,
the synchrotron X-ray luminosity {\em does not depend} 
on the strength of magnetic field.  
In this regime an equilibrium is quickly established between the electron
injection and synchrotron losses. This implies that the synchrotron radiation
saturates to the maximum possible rate determined by the electron 
injection rate, $L_\mathrm{synch} \simeq \dot{W}_\mathrm{e}$.
%%Yet the strength of the B-field may have an impact on the position of the
%%synchrotron cutoff ($\propto B$). However, in the case of diffusive shock 
%%acceleration that proceeds in the Bohm diffusion limit, this dependence 
%%also disappears as long as the acceleration and radiation regions coincide.
Therefore, below we assume 
that the filamentary structure is associated, first of all,
with inhomogeneous spatial injection of electrons.  
In general, since the propagation speed in different parts of the rim could be 
very different depending on the magnetic field and the developed turbulence, 
we should expect significant spatial variation of the density of particles 
even in those cases when they are injected into the rim almost homogeneously.  
The acceleration of electrons in selective regions of the rim  would make 
the X-ray synchrotron image of the rim even more irregular and clumpy. 

In this context,  it is challenging to identify the X-ray  filaments as sites
of effective  electron acceleration to TeV energies. It is likely that 
the filaments  are bright not only due to enhanced 
magnetic fields, but also due to local concentrations  of 
the multi-TeV electrons being directly  accelerated there.
To our knowledge, the filamentary structure is a feature which has not been
discussed (dismissed ?) in the relatively well developed concept of  
diffusive shock acceleration in SNRs, although 
perhaps this could be attributed to the inhomogeneity of the 
ambient medium.
%%(e.g. {\bf McKee 1982})
Also,  it is  interesting to
note in this regard,  that an appearance of ``hot spots'' or ``discharge zones'' 
has been qualitatively predicted by Malkov (\cite{malkov97}) who related 
such  features to the drop of the critical injection caused by changes in the 
flow structure.   Thus the very fact of existence of nonthermal 
filamentary structures of the shell of this SNR should not be
interpreted as contradiction to  the diffusive shock acceleration model.
The unusually small spatial scales of these structures perhaps also 
could be accommodated by this theory assuming very large ambient 
magnetic field significantly exceeding $10 \ \rm \mu G$.  However,
as discussed above,
any assumption of the magnetic field cannot explain the position of 
the synchrotron cutoffs beyond 10 keV, unless we assume more effective 
acceleration mode (e.g. Jokipii \cite{jokipii87}).  The question of the nature 
of the mechanism of particle acceleration is beyond the framework of this paper, therefore below we will adopt {\em a priory} 
an acceleration spectrum extending well beyond 100 TeV 
at the presence of relatively large field -- 
a condition dictated directly by the X-ray observations --
and explore the spectral and spatial features of synchrotron radiation 
associated with the propagation effects of electrons in the rim.  

The small lateral dimensions of the 
compact structures in the filaments allow effective escape of accelerated 
electrons, thus the main part of electrons would eventually 
reside outside their production sites. Therefore,  it is not a surprise that 
the total  X-ray flux from the NW rim is contributed mainly 
by the plateau region,  although the filaments  look much brighter.    
  
Given the fast synchrotron cooling of X-radiating electrons
compared to the source age and the compactness of the accelerators, 
any appropriate treatment of the problem 
should be \emph{time-dependent} and should include \emph{propagation effects}. 
More specifically,   for the synchrotron cooling time  
$\tau_{\mathrm{syn}} \approx
760\, (\varepsilon/5\,\mathrm{keV})^{-1/2} (B/10\,\mu\mathrm{G})^{-3/2}$~yr, 
the spectral steepening of X-ray emitting electrons becomes unavoidable 
even for the minimum possible age of the source, $\tau_0 \sim 1000$~yr.
The energy-dependent escape on timescales 
$\tau_{\mathrm{dif}} \simeq 
\Delta R^2/2\mathcal{D}_\mathrm{B} \simeq 
1500\, d_6^2 \eta^{-1}(E/10\ \mathrm{TeV})^{-1} (B/10\,\mu \mathrm{G})
 \ \mathrm{yr}$,
is another important factor for  modification  of the electron 
spectrum in thin filaments.       
Here  $\Delta R = 0.58\,d_6$ pc is the typical thickness of filaments,  
and $\mathcal{D}_\mathrm{B}(E)=\eta r_{\mathrm{g}}c/3$ is 
the Bohm diffusion coefficient with the gyrofactor as a free parameter. 
In addition to the diffusive escape, we should take into account also   
the  convective escape   that occurs on timescales 
$\tau_{\mathrm{con}} \sim \rho \Delta R /V$
where $\rho$ is the shock-compression ratio.
The convection is  an important effect for delivery of
low-energy particles from filaments to the surrounding plateau region.

To simplify the problem, and to clarify the basic  relations,
here we adopt  a  ``two-zone'' model  applied earlier to SN 1006
(Aharonian \& Atoyan \cite{aharonian99}). This model takes into account
the effects related to the diffusive and convective escape of electrons 
from one homogeneous region (zone 1: acceleration site, i.e. filaments)  
to another (zone 2: the rest of the rim, i.e. plateau, where electrons coming from 
zone 1 are accumulated without escape) with essentially different 
physical parameters  such as magnetic field, diffusion 
coefficient, {\em etc.},  and gives time-dependent 
solutions for energy distributions of  electrons   in both zones.  
In the \emph{Chandra} image of the NW rim,
the classification of the filament/hot-spot (zone 1)  and plateau (zone 2) regions 
is based on the surface brightness.
The overall (i.e. integrated over the volumes)  fluxes in the 2--10 keV band 
are estimated to be 
$1.1 \times 10^{-11}\, \mathrm{erg}\ \mathrm{cm}^{-2}\, \mathrm{s}^{-1}$ and
$4.2 \times 10^{-11}\, \mathrm{erg}\ \mathrm{cm}^{-2}\, \mathrm{s}^{-1}$
for zone 1 and 2, respectively. 
Figure\,\ref{fig:multiband} shows the X-ray spectra from zone 1 and 2
characterized by a power-law function with a photon index $\Gamma = 2.2$.
Below, we try to reproduce the observed fluxes and spectra of 
synchrotron X-rays from the filament and plateau, simultaneously, 
in the framework of the ``two-zone'' model, in order to testify the 
``filament/hot-spot = accelerator'' scenario.

For the case of an old ($\tau_0 =10\,000$ yr) and distant  ($d = 6$ kpc) SNR,  
we found that a satisfactory  fit is possible assuming  strong and similar 
magnetic fields  in the filaments and plateau, 
$B_{\mathrm{fil}} = B_{\mathrm{pla}} = 50\,\mu\mathrm{G}$.
Since the strong magnetic field implies fast synchrotron cooling,  
in order to get large X-ray fluxes from the plateau
we should require an adequately fast escape  of electrons 
from the filaments, namely $\tau_{\mathrm{dif}} < \tau_{\mathrm{syn}}$, or 
$\eta > 16 \,(B_{\mathrm{fil}}/50\,\mu\mathrm{G})^3 d_6^2$. 
In Fig.\,\ref{fig:multiband}a we assume $\eta = 83$ and  
$\tau_{\mathrm{con}}=1000$ yr. 
The X-ray slope of the filament is explained by 
synchrotron radiation from the 
\emph{diffusive-escape-steepened} distribution of high-energy electrons,
while that of the diffuse plateau -- from the 
\emph{synchrotron-loss-steepened} distribution. The derived 
parameters lead to quite similar X-ray spectra for two zones.
The total energy of electrons accumulated   in the plateau,
$W_{\mathrm{pla}} = 4.2 \times 10^{47}\ \mathrm{erg}$,
appears one order of magnitude larger than that in the filaments.
On the other hand, 
if the SNR is much younger ($\tau_0 = 1000$ yr) and closer ($d = 2$ kpc),
 a good fit is possible 
assuming that the magnetic field is stronger in the filaments, namely
$B_{\mathrm{fil}} = 20\,\mu\mathrm{G}$, but weaker in the plateau,
$B_{\mathrm{pla}} = 6\,\mu\mathrm{G}$.
Since in this case the diffusive escape is very fast 
($\tau_{\mathrm{dif}} \propto \Delta R^2  \propto d^2$), 
we must assume Bohm diffusion regime, $\eta =1$, 
in  order to keep   electrons within the filaments. We also assume 
$\tau_{\mathrm{con}}=500$ yr. 
Due to the smaller magnetic field in the plateau region 
and the small $\tau_0/\tau_{\mathrm{con}}$ ratio,
the radio emissivity of  the plateau is considerably suppressed. 
The radio flux coming from the NW rim region 
plotted in Fig.\,\ref{fig:multiband} should not exclude this model 
because the population of radio-emitting electrons and that of 
X-ray-emitting electrons could be very different;
the high quality data of the radio fluxes would be needed prior to
definitive conclusions.

Modeling of the observed synchrotron X-ray emission
is crucial for  predictions of  the spatial and spectral 
distributions of the associated IC $\gamma$-rays produced by the same 
electrons upscattering 2.7 K CMBR.  
In Fig.\,\ref{fig:multiband} we  present the IC spectra  
from the filaments and the plateau. In both cases,
we found that it is quite difficult to 
reproduce the flux and spectral shape of TeV $\gamma$-rays  
reported by the  CANGAROO collaboration, in the synchrotron-IC model.

%_____________________________________________________________
%                 A figure as large as the width of the column
%-------------------------------------------------------------
   \begin{figure}
   \centering
       \includegraphics[width=8cm]{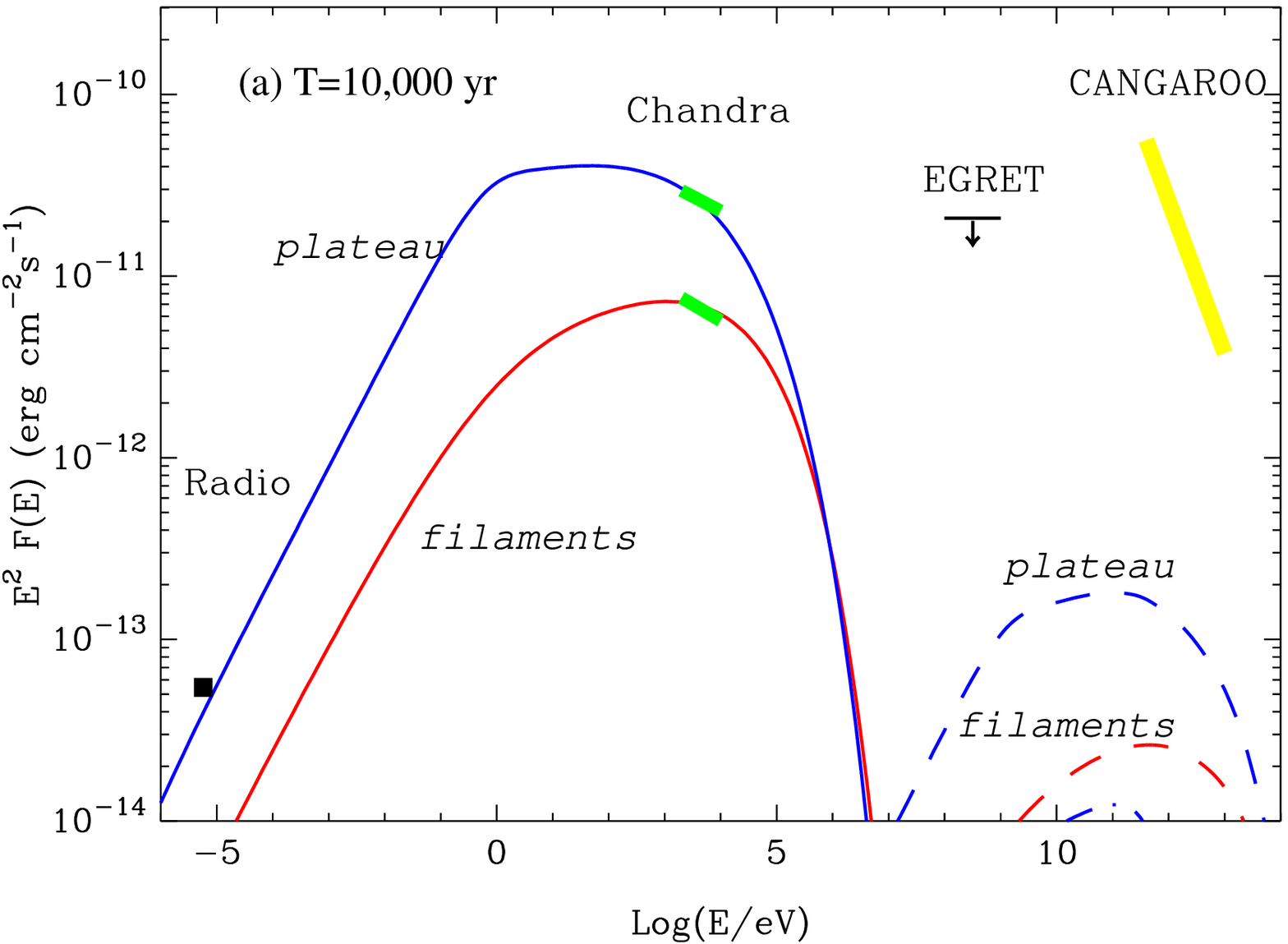}
       \includegraphics[width=8cm]{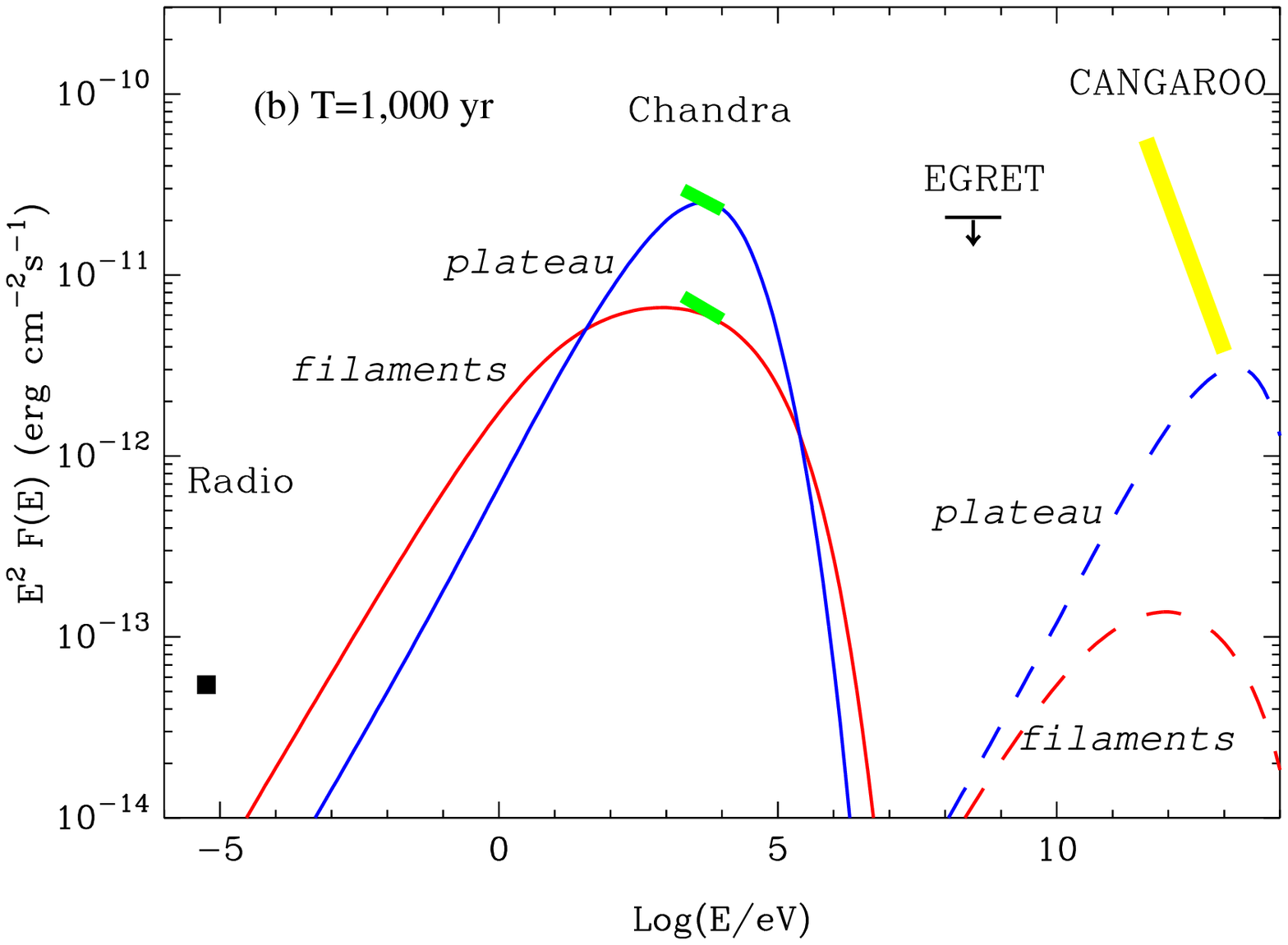}
      \caption{Multiwavelength synchrotron (\emph{solid  lines})
and IC (\emph{dashed lines}) spectra of the filaments and the plateau regions 
calculated within the ``two-zone'' model,
assuming that the electron acceleration takes place in the filaments.
The ATCA flux of 4~Jy at 1.36~GHz (\emph{squares}) is from 
Ellison et al.\ (\cite{ellison}).
The X-ray spectra from the filaments and the plateau region 
are from this work, 
and the TeV $\gamma$-ray flux is  from the CANGAROO observations
(Enomoto et al.\ \cite{enomoto}).
The EGRET upper limit corresponds to the flux of 3EG J1714$-$3857 
(Butt et al.\ \cite{butt01}).
The following parameter sets have been used in calculations:
{\bf a)} 
$\tau_0 =10\,000$ yr and $d = 6$ kpc. $\tau_{\mathrm{con}}=1000$ yr.
Acceleration spectrum of electrons  with the spectral index $s=1.84$,  
exponential cutoff at $E_0 = 125$ TeV, and acceleration rate  
$L = 2.8 \times 10^{36}\ \mathrm{erg}\ \mathrm{s}^{-1}$. 
$B_{\mathrm{fil}} = B_{\mathrm{pla}} = 50\,\mu\mathrm{G}$ and $\eta = 83$.
{\bf b)}
$\tau_0 =1000$ yr and $d = 2$ kpc,  $\tau_{\mathrm{con}}=500$ yr, 
$s=1.95$, $E_0 = 200$ TeV, and 
$L = 1.6 \times 10^{37}\ \mathrm{erg}\ \mathrm{s}^{-1}$. 
$B_{\mathrm{fil}} = 20\,\mu\mathrm{G}$,
$B_{\mathrm{pla}} = 6\,\mu\mathrm{G}$ and $\eta = 1$.
}
         \label{fig:multiband}
   \end{figure}

It should be noticed that the time-dependent treatment
of the problem requires very effective acceleration of electrons. 
The exponential cutoffs in the {\em acceleration spectrum}
are assumed at 125 TeV (case {\bf a}) and 200 TeV  (case {\bf b})
in order to fit the X-ray data. As it follows from Eq.\,(\ref{Emax}) 
these assumptions   hardly could be  accommodated by  the standard
diffusive shock acceleration model. 

Finally, we would like to make a  short comment  
concerning the possible association of the X-ray void
with one of the three molecular clouds (cloud~C, see Fig.\,\ref{fig:fov}) 
reported  by Slane et al.\ (\cite{slane}) 
in the proximity of RX~J1713.7$-$3946. 
The density of this cloud is estimated 
from CO observations (Bronfman et al.\ \cite{bronfman}),
$n \sim 400\,d_6^{-1}\ \mathrm{cm}^{-3}$.
If cloud~C  has physical association with   
RX~J1713.7$-$3946, it could be an interesting site for the production 
of very high energy $\gamma$-rays through interactions of 
electrons and protons with the ambient dense gas.
In order to explain the bulk of the reported TeV 
luminosity (Enomoto et al.\ 2002),  
$L_{\gamma} \sim 4 \times 10^{35}\,d_6^2\ \mathrm{erg}\ \mathrm{s}^{-1}$,
by $\pi^0$-decay $\gamma$-rays, the total energy of  protons 
in this cloud should be close to $10^{50} d_6^2 \ \rm erg$.
This exceeds, by 2 or 3 orders of magnitude,  the 
total energy released in relativistic electrons,   assuming that the 
particle accelerator has been operating during the age of the source:
$W_{\rm e}=\tau_0 L_{\rm e} \simeq 8.8 \times 10^{47}$ and 
$5 \times 10^{47}$ erg for $d=6$ and 2 kpc, respectively 
(see Fig.\,\ref{fig:multiband}).
Thus,  the hypothesis of the $\pi^0$-decay origin of the reported 
TeV flux would require  a very large 
proton-to-electron acceleration ratio,  significantly exceeding
the canonical ``100:1'' value,  given that only a fraction  of the protons 
accelerated by the supernova shock could be captured by cloud C.   

\section{Conclusions}

The \emph{Chandra} image has revealed that the synchrotron X-ray emission from
the northwestern rim of SNR RX~J1713.7$-$3946 has remarkable 
fine-structure: the complex network of synchrotron
X-ray filaments surrounded by fainter diffuse plateau, and a 
dark region with a circular shape.
By examining individual spectra, we found that 
despite significant brightness variations, 
the spectral shapes of the X-ray spectra everywhere in this region
are more or less similar, being well fitted with a power-law model 
of photon index $\Gamma \simeq$ 2.3.

The observed hard power law requires rather high synchrotron cutoff 
frequency (energy), 
set by the condition ``acceleration rate = synchrotron loss rate'',
which is most likely to be $\gtrsim10$ keV taking account of the effects of 
spectral steepening due to synchrotron losses.
We need unreasonably high shock speed exceeding 5000 km s$^{-1}$
to explain such a high cutoff energy within the standard formalism of the
diffusive shock acceleration model.
A possible solution to this difficulty could be obtained if one assumes 
low magnetic field in the acceleration sites, and high magnetic field
in the emission regions.
Otherwise we should invoke faster, as yet unknown, electron 
acceleration mechanism.

We testify the scenario that the filaments/hot-spots are the acceleration 
sites of bulk of multi-TeV electrons without specifying acceleration mechanism,
by adopting a ``two-zone'' model. 
Then we found that our time-dependent treatment 
including the effects of fast energy losses and diffusive escape of electrons is 
capable of  accounting for the fluxes and spectral shapes of X-ray emissions from
both the filament and plateau regions.
We note that the TeV spectrum reported by the CANGAROO collaboration cannot 
be readily explained by the IC scattering of the CMBR photons 
off X-ray-emitting multi-TeV electrons.
Higher quality data from the radio, X-ray and $\gamma$-ray bands should 
help us to draw more definitive conclusions concerning the nature of acceleration 
mechanisms and the origin of the observed TeV $\gamma$-rays.

\begin{acknowledgements}
We thank the referee, B.~Aschenbach for very useful comments which
helped to improve the paper, 
as well as G.~Rowell and Y.~Butt for helpful comments.
Y.U. is supported by the Research Fellowships of
the Japan Society for the Promotion of Science for Young Scientists.
\end{acknowledgements}

\end{document}